\documentclass{ragtime} 

\title[Spinning test particles in equatorial plane of KdS spacetimes]%
      {Low angular momentum\footnote{see Introduction for specification of what we mean by low angular momentum} accretion flow model 
       of Sgr A* activity}

\author[B. Czerny,  
        M. Mo\' scibrodzka,     
        D. Proga,
        T. Das,
        and A. Siemiginowska]
       {Bo\. zena Czerny\at{1,a} 
        Monika Mo\' scibrodzka\at{1} 
        Daniel Proga\at{2}
        Tapas K. Das\at[]{3}
        and Aneta Siemiginowska\at[]{4}\\
        \ins{1}Copernicus Astronomical Center,
        Bartycka~18, 00\,710 Warsaw,
        Poland\\
        \ins{2}Department of Physics, University of Nevada, 
        Las Vegas, NV 89154, USA\\
        \ins{3}Harish Chandra Research Institute, Allahabad-211 019, India
        \\
        \ins{4}Harvard-Smithsonian Center for Astrophysics, 60 Garden Street, 
        Cambridge MA 02138, USA\\
        \ins{a}\Email{bcz@camk.edu.pl}} 

\coentry{B. Czerny et al.}%
       {Low angular momentum accretion flow model\par
        of Sgr A* activity}

\begin{document}

\begin{abstract}
 Sgr A* is a source of strongly variable emission in several energy bands. It is 
generally agreed that this emission comes from the material surrounding the black 
hole which is either falling in or flowing out. The activity must be driven by
accretion but the character of accretion
flow in this object is an open question. We suggest that the inflow is dominated
by the relatively low angular momentum material originating in one of the nearby 
group of stars. Such material flows in directly towards the black
hole up to the distance of order of ten Schwarzschild radii or less, where it
hits the angular momentum barrier which leads naturally to a flow variability.
We study both the analytical and the numerical solutions for the flow dynamics, 
and we analyze the radiation spectra in both cases using the Monte Carlo code 
to simulate the synchrotron, bremsstrahlung and the Compton scattering. Our model
roughly reproduces the broad band spectrum of Sgr A* and its variability if we
allow for a small fraction of energy to be converted to non-thermal population of 
electrons. It is also consistent (for a range of viewing angles) with the strong 
constraints on the amount of circumnuclear material imposed by the measurements of 
the Faraday rotation.
\end{abstract}

\begin{keywords}
Sgr A*~-- black hole~-- accretion~--
radiation spectra
\end{keywords}

\section{Introduction}\label{intro}

The character of the accretion flow onto a black hole depends on the initial 
angular momentum of the material. This angular momentum is specified by the
outer boundary conditions which depend on the relative motion of the donor with
respect to the black hole. This angular momentum corresponds to a certain circularization
radius, i.e. the radius where this angular momentum is equal to the local Keplerian
value. In binary systems the material comes from the secondary
star and in general is possess high angular momentum due to the orbital motion. 
In Low mass X-ray binaries the flow proceeds through an inner Lagrange point and the
circularization radius is a significant fraction of a Roche radius around a 
black hole, of order of $10^4 R_g$ ($R_g = GM/c$). In high mass X-ray binaries the
accretion flow comes from the intercepted focused wind, so the circularization radius
is smaller but still large, of order of $10^3 R_g$. In such case the inflowing material 
form an accretion disk around a black hole and the inflow proceeds due to the angular 
momentum transfer. Apart from the outermost region and the region close to the ISCO 
(innermost stable circular orbit), the distribution of the angular momentum is 
relatively smooth and not much different from the Keplerian law. The exact departures 
from Keplerian motion depends on the disk temperature (or more exactly, on the pressure 
distribution).

In active galactic nuclei (AGN) 
the source of material is less specified. The material comes either from the stars 
(in the form of stellar winds) or from the gaseous phase of the galactic material. 
Bright AGN (quasars, Seyfert 1 galaxies) 
show the presence of accretion disks similar to the disks in binary systems so we can
conclude that the angular momentum reaching the galactic center is high. In sources 
showing water maser activity we observe the outer parts of the disk directly, and in 
most sources the motion of the disk material is Keplerian. However, in weakly active
galaxies like Sgr A* or giant elliptical galaxies we see no direct evidence of a disk.
In Sgr A* the presence of the cold disk is actually excluded by the lack of eclipses of
the stars which move very close to the central black hole and are systematically 
monitored since several years. 

Since in weakly active galaxies there are no direct observational arguments for any 
value of the angular momentum
of the donated material and the location of material sources, three types of models are
being considered:  
\begin{itemize}
\item high angular momentum flow, with circularization radius of order of hundreds-thousands of $R_g$
\item low angular momentum flow, with circularization radius of order of a few  $R_g$
\item spherical and quasi-spherical accretion, without angular momentum barrier.
\end{itemize} 

The high angular momentum flow solutions for weakly active galaxies generally belong to
ADAF (advection dominated accretion flow) family \citep{1977ApJ...214..840I,1994ApJ...428L..13N}, with possibly additional effects like
outflows \citep{1999MNRAS.303L...1B} and convection. In this case the flow is not exactly Keplerian since
the pressure gradients are important, but the local ratio of the angular momentum to the
Keplerian angular momentum in most part of the flow is not wildly different from unity,
and the angular momentum transfer (through viscosity) or angular momentum loss (through 
magnetic wind) at all radii is essential. Stationary solutions usually exist, and 
asymptotically the density of the flow approaches zero at infinity.

In spherical and quasi-spherical flow there is no angular momentum barrier so the loss
of angular momentum is not the necessary condition for the accretion to occur. Examples
of such solutions are: purely spherical Bondi flow or flows where the angular momentum
density is below the minimum angular momentum at the circular orbit around a black hole 
which is 
given by
\begin{equation}
l_{min} = 3 \sqrt{3} GM/c
\end{equation} 
in case of Schwarzschild black hole; more general formula for a Kerr black hole can be
found in \citep{1972ApJ...178..347B}.
In Bondi solution \citep{1952MNRAS.112..195B,2003ApJ...591..891B} the outer boundary condition are specified by the density and the
temperature of the uniform medium surrounding black hole at large distances. The flow
velocity is zero at infinity, the inflow becomes transonic at the Bondi radius, and the
supersonic flow reaches the black hole horizon. The Bondi radius depends significantly
on the gas properties (e.g. politropic index; Bondi radius is of order of thousands of
$R_g$ for relativistic flow with $\gamma = 4/3$ but is approaches zero if 
$\gamma \rightarrow 5/3$, typical for perfect fluid non-relativistic solution), 
but the accretion rate is 
much less 
sensitive to those assumptions.Purely Bondi flow has generally very low radiative efficiency so it cannot reproduce the observed luminosity in most weakly active galaxies \citep{2006A&A...450...93M}.
If the accreting material at the outer boundary condition has certain angular momentum $l < l_{min}$, the dynamics of the flow is slightly modified in comparison with Bondi flow 
and the flow is not spherically symmetric any more but the stationary solution for the
flow always exists.

The intermediate case of low angular momentum the situation is the most complex as
initially the flow behaves as the Bondi flow but close to the black hole the flow starts
suddenly to feel the angular momentum barrier \citep{1981ApJ...246..314A}. In this case analytical stationary 
solutions frequently do not exist. In numerical solutions the flow is variable and does
not reach a stationary solution in the computing time. If the angular momentum of the
donated material is also a subject of changes (e.g. the result of the stellar motion),
a truly stationary solution indeed can never be reached for physical reasons.

In the case of Sgr A* the available spatial resolution is the highest and we can have
the best insight into the sources of material \citep{2007IAUS..238..173G}. Therefore, in the present paper we
concentrate specifically on this source and we argue that the low angular momentum
flow is an interesting and promising option for the flow description.

\section{Sgr A* surrounding and the source of material}\label{sect:surrounding}

At the central parts of our Galaxy there are a few stellar populations and each of them
provides stellar winds. The closest O/B stars (at distances of a small fraction of a parsec) used for the mass measurements have moderate winds; much stronger winds come from more evolved stars being at distances up to a few parsecs. If a single star dominates as a donor star, and the wind velocity is larger than the stellar orbital velocity, there is a zero angular momentum line joining the donor star and the black hole, so the net angular momentum flowing in is likely to be low \citep{2004MNRAS.350..725L,2006MNRAS.370..219M}. Estimates of the ram pressure indicate that a single source, IRS 13 E (the compact group of Wolf-Rayet stars) with the strongest wind (e.g. \citet{2004ApJ...604..662R}) indeed dominates independently from the relatively large distance from the central black hole \citep{2006MNRAS.370..219M}.  

Although the ram pressure argument strongly depends on the adopted wind outflow rate and wind velocity, the additional argument against the dominance of the nearby young O/B stars comes from the lack of obvious correlation between the activity level and stellar passages. On the other hand, un-modulated flow can come from stars forming a mysterious ring-like structure at a distance of a parsec scale \citep{2006ApJ...643.1011P}. In this case the material might have very large angular momentum, and even a cold disk may form, as postulated by \citet{2003MNRAS.343L..15N}. However, the periodicity seemed to be seen in the NIR and X-ray flares \citep{2003Natur.425..934G,2006JPhCS..54..420B} and the absence of eclipses does not seem to support cold disk scenario. The motion of IRS 13E differs from that of other stars and seems to be significantly eccentric. Detailed studies of the observational consequences of all scenarios are necessary to solve the issue. 

\section{Analytical solutions for the dynamics}\label{KdS}

Analytical solutions can be obtained if the description of the flow is simplified: 
outflow and viscosity are neglected, the flow is politropic and the gravity is described by pseudo-Newtonian potential. The problem was studied in numerous papers following the idea of \citet{1981ApJ...246..314A}. 

\subsection{Transonic solutions with and without shocks}

The existence and the character of solution depends critically on the interplay of the
solution parameters: politropic index, energy density and angular momentum density, e.g. \citet{2003ApJ...592.1078D}.
The equations usually show the presence of three critical points: the outer one is direct generalization of the Bondi radius, the inner one describes the expected transition to the supersonic flow below the ISCO and the intermediate point is unphysical (in a sense it is not of the saddle type and the flow cannot pass there smoothly from subsonic to supersonic solution). Usually the flow passes from subsonic flow at infinity to supersonic flow close to horizon either at the outer {\it or } at the inner critical point. For a certain range of the parameters, it is also possible to find a second solution, with a shock (satisfying the standard Rankine-Hugoniot conditions) located  between the outer and the inner sonic point. Such a shock {\it may}, but {\it not must}, form. 

\begin{figure}[t]
\begin{center}
\includegraphics[width=\linewidth,height=0.6\linewidth]{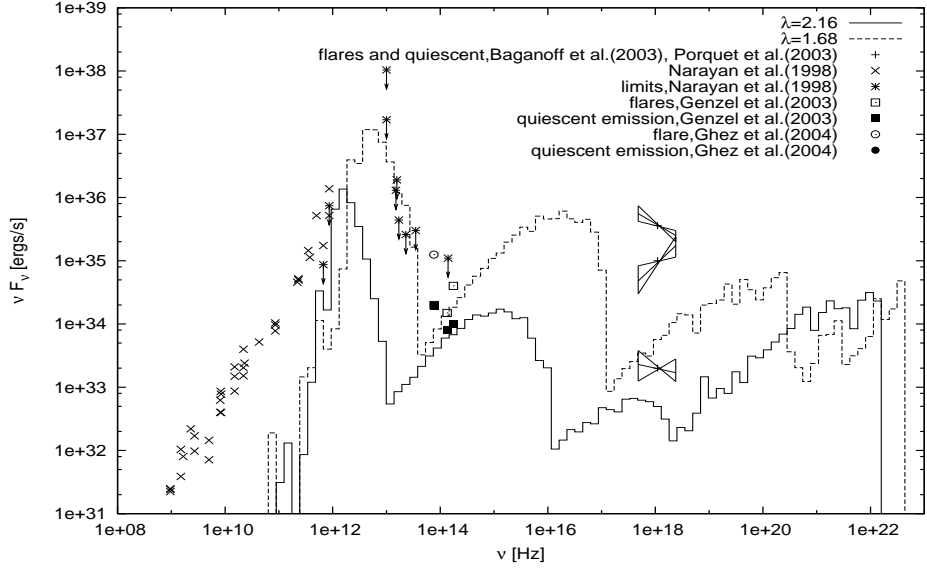}
\end{center}
\caption{\label{Fig:an_spe}The two exemplary spectra models of SGR A* for two values of the angular momentum density in analytical model. One of the solutions extends only down to $20 R_g$ and correspond naturally to the weakly variable part of the flow. Thermal distribution of electrons was assumed. Dots mark the data points or upper limits in radio and
NIR, and three representative levels of radio emission from Chandra are marked as power laws with slope errors. }
\end{figure}

The eventual shock development is likely to be related to the past state of the accretion flow. For a fixed value of politropic index and energy, there is a specific value of the angular momentum density at which the solution changes from transonic at outer to transonic at inner critical points. This change of flow properties is dramatic, and therefore the flow slowly crossing this angular momentum border is likely to develop a shock instead of following a new shock-less solution since this allows for a slow and continuous change in the flow properties (shock is initially weak). The discussion of this issue will be presented elsewhere (Das, Czerny \& Mo\' scibrodzka, in preparation).

For large enough angular momentum there is no analytical solution which extends down to the black hole even if a shock is allowed, and in this case the dynamics can be described analytically only e.g. down to a few a a few tens of $R_g$ \citep{2006MNRAS.370..219M}. This part of the flow is likely to be stationary while the inner part must form a kind of unstable ring.

\subsection{Exemplary spectra}

The radiation spectra were calculated with the code described in \citet{2006A&A...450...93M} and later generalized to non-spherical distribution of the inflowing material \citep{2006MNRAS.370..219M}. The emission processes taken into account include synchrotron radiation, bremsstrahlung and Compton scattering. 

Two examples of the spectra are shown in Fig.~\ref{Fig:an_spe}. The spectra reproduce the NIR peak although are short of the data points at long wavelengths. The level of the continuous emission in X-rays is reproduced by the higher angular momentum model, so we can speculate that the emission of the innermost unstable ring, not described by a stationary solution, may account for the strongly variable part of the X-ray flux. 

\section{MHD simulations of the flow dynamics}

\subsection{Flow description}

Exploratory phase of the use of MHD simulations for modeling Sgr A* time-dependent
spectra made use of the simulations which were performed by \citet{2003ApJ...592..767P}.
The setup assumed the almost Bondi flow. The input of material was set at 1.2 $R_B$
(where $R_B$ is the Bondi radius), which in turn was equal to $2 \times 10^3 R_g$,
and the angular momentum of the new material was $4R_g c$ at the equatorial plane.    
The time-dependent computations were performed for a few dynamical timescales at the
outer radius which gave the time-dependent distribution of the energy density, 
velocity and the magnetic field. The flow was strongly time-dependent, with large scale 
fluctuations seen at the end of the computer run as well. The inflow was accompanied by 
a significant outflow. The computations never fully reached the exact stationarity condition in a sense that the inflow rate was not quite balanced by the outflow and accretion rate. However, in reality the flow also may not have time to reach equilibrium since the donor star moves (long term effects) and the Wolf-Rayet stellar winds show considerable clumpiness and overall variability (short timescale effects).

Time-dependent electron temperature distribution was obtained by assuming several channels of energy transmission: (i) Coulomb coupling between the ions and electrons, (ii) 
direct heating of electrons due to compression (iii) electron radiative cooling (iv) electron advection. Optionally, we also allowed a fraction of energy to be in a form of non-thermal electrons. In order to simplify the computations, at each moment the flow was assumed to be stationary i.e. time-dependent solutions for the flow dynamics were treated as frozen frames. For a detailed description, see \citet{2007arXiv0707.1403M}. 

\subsection{Spectral variability}

\begin{figure}[t]
\begin{center}
\includegraphics[width=\linewidth,height=0.6\linewidth]{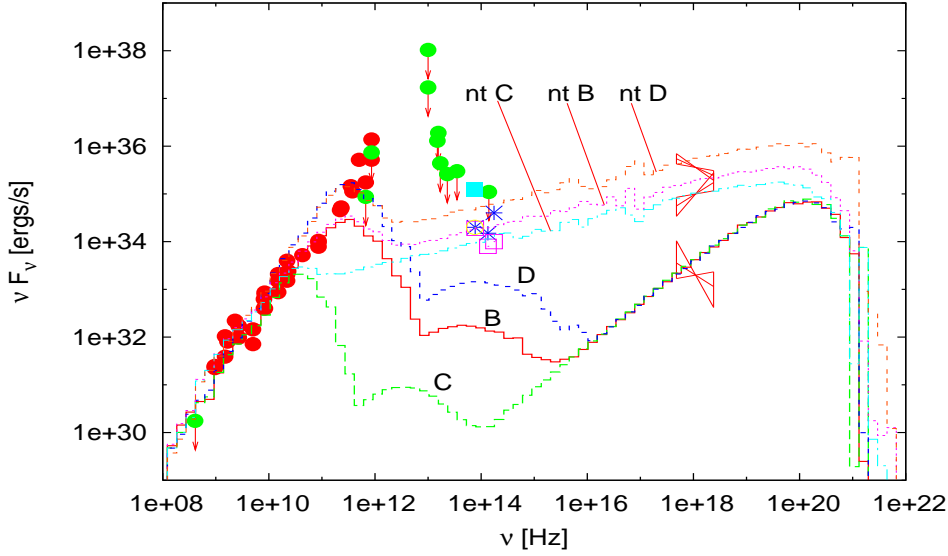}
\end{center}
\caption{\label{Fig:1}MHD simulations: the exemplary broad band spectra of SGR A* for four different time
moments labeled as A,B,C and D. Dots mark the data points or upper limits in radio and
NIR, and three representative levels of radio emission from Chandra are marked as power laws with slope errors. Lower lines represent solutions with only thermal electrons, upper lines show solutions with a fraction of energy in a population of non-thermal electrons.}
\end{figure}

Flow variability was reflected in variability of the broad band spectra. Exemplary states
are shown in Fig.~\ref{Fig:1}. We see that the pure thermal electron distribution cannot 
represent the X-ray variability. The variable emission comes from the inner region of the flow in the form of synchrotron emission, and for thermal electrons this component does not extend to X-rays. Bremsstrahlung emission, and Compton scattered emission comes from more extended region where variability is weaker and/or smeared. Therefore, the presence of non-thermal population is essential.

The amplitude of the variability in the timescales recorded in MHD simulations (a day) was very large, over an order of magnitude so the variability in the overall accretion rate (one order of magnitude) is additionally enhanced at some wavelengths (particularly at NIR) by the spectral effects.

The relation between the dynamics and the spectra is rather complex as it is strongly non-linear. There is no one-to-one correspondence between the accretion state and the predicted spectrum. For example, two states with similar accretion rates at the inner edge can have different spectra, or two states with different accretion rates can have similar spectra. 

The specific issue is the description of the outer parts of the flow. Bremsstrahlung dominates there, and the measured flux depends on the size of the emitting region. Since the spatial resolution available in X-ray band is low in comparison with the modeled region (even in Sgr A*!) this is an additional problem in comparing the models to the data. 

\subsection{Faraday rotation measure}
\begin{figure}[t]
\begin{center}
\includegraphics[width=\linewidth,height=0.6\linewidth]{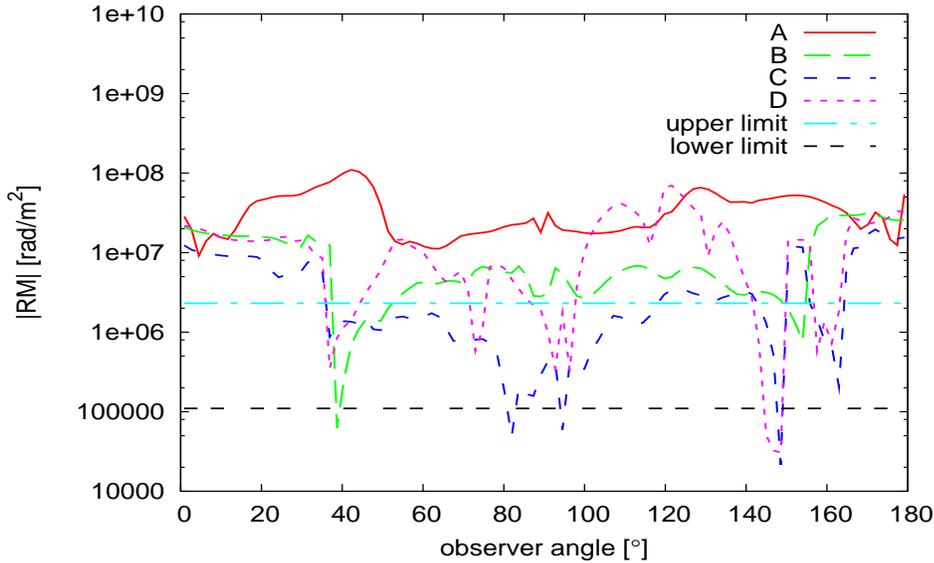}
\end{center}
\caption{\label{Fig:rot} The determination of the Faraday rotation measure from the model, as a function of the viewing angle of an observer, for the same four different time
moments labeled as A,B,C and D. Horizontal lines show the upper and the lower limit from the measurements of \citet{2006ApJ...640..308M}.}
\end{figure}

The strongest observational constraint for the models comes from the estimates of the Faraday rotation measure towards Sgr A*. The observed polarization and the change of its angle indicate very low density plasma along the line of sight to the source. Our MHD model allows to calculate the integrated expected rotation measure as a function of the line of sight. The result is shown in Fig.~\ref{Fig:rot}. For some inclination angles the model is consistent with the observational limits.

\section{Recent developments}

\subsection{GR effects}

\begin{figure}[t]
\begin{center}
\includegraphics[width=0.5\linewidth,height=0.5\linewidth, angle=-90]{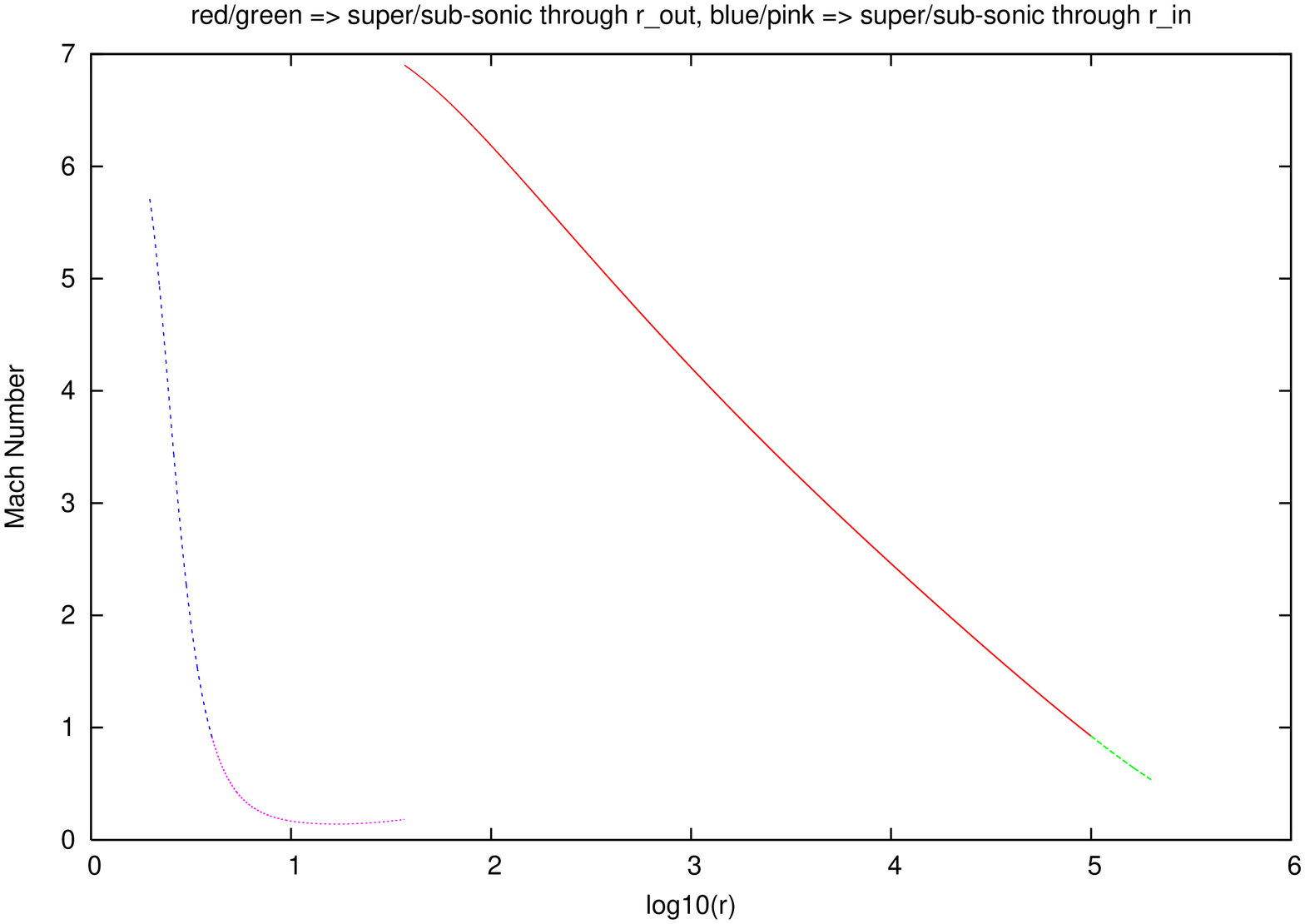}
\includegraphics[width=0.5\linewidth,height=0.5\linewidth]{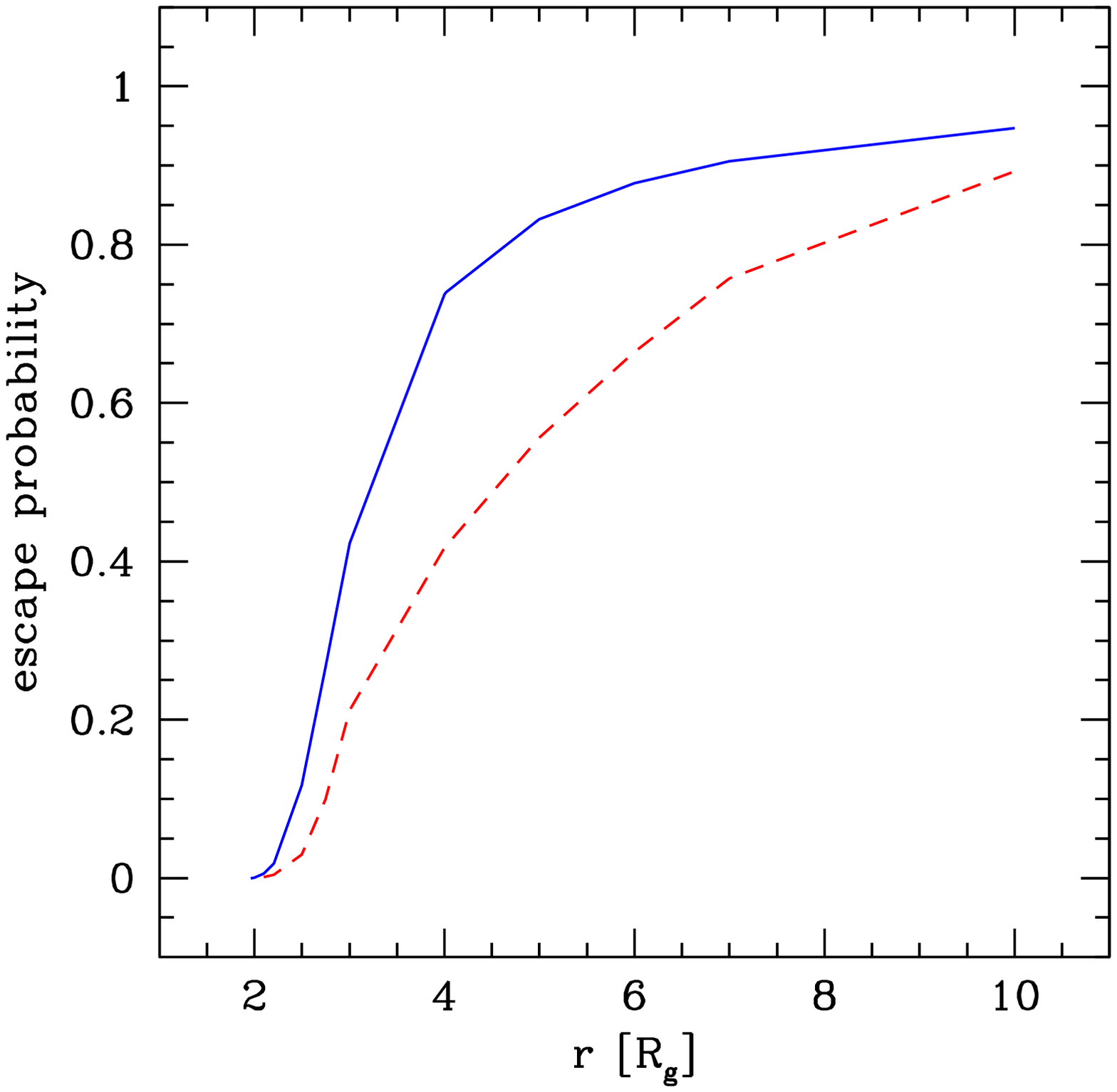}
\end{center}
\caption{\label{Fig:Mach}The radial dependence of the Mach number for a solution with the shock and the corresponding escape probability; Kerr parameter $a=0.3$. Dashed line in the lower panel shows the escape probability for a solution without a shock, with much larger radial velocity of the emitter in the innermost part of the flow.}
\end{figure}

The computations presented in previous sections did not include the effects of General 
Relativity properly --- either purely Newtonian approach was used (with GR effects mimicked by the flow cut-off at $6 R_g$) or pseudo-Newtonian potential was adopted. Since the black hole in Sgr A* is likely to be rotating, more appropriate approach would be useful, or at least some estimates must be performed of an error to the spectrum due to the negligence of GR effects.

Analytical solutions for the low angular momentum flow in the Kerr metric are well known \citep{2004ApJ...613L..49B} and can be used to determine the flow dynamics. Spectral computations using the
Monte Carlo code require computing millions of photons so ray tracing for each of them is too time consuming if a number of models is to be calculated. However, we can 
calculate the escape probability of a photon as a function of radius, for a given Kerr parameter and we can test whether it depends significantly on the motion of the emitter (i.e. the solution for the flow dynamics). 

The computations are done in a standard way (see e.g. \citet{2005ragt.meet..143S} and the references therein) by integrating the photon paths for 100 000 photons emitted isotropically within the frame of the emitter at a given radius, and the calculations are performed for several radial points. 

This work is still in progress, but in Fig.~\ref{Fig:Mach} we show an example for a specific
dynamical solution with a shock. For a comparison, in the lower panel we also show the result for the solution without a shock (dashed line). We see that the material effective emissivity is much higher in the case with shock, even without the shock emission included. Therefore, the GR effects should be estimated separately for each dynamical solution.

\subsection{New dynamical simulations}

Also the dynamical MHD simulations used in the previous study was not fully 
satisfactory from the point of view of Sgr A* modeling. First of all, the 
results of the simulations were not recorded densely enough in time to allow us to follow
as fast time variability as 17 minutes seen in QPO-like events. New simulations, better suited for the center of our Galaxy, are currently being performed (Mo\' scibrodzka
\& Proga, in preparation). The preliminary results show that indeed the shortest
timescale variability is seen in the dynamical simulations.

\section{Conclusions}\label{conclus}

Low angular momentum accretion flow is a promising scenario for the accretion onto Sgr A* due to its natural variability pattern. The flow is slightly more energetically efficient than the purely spherical Bondi flow and can reproduce both the required level of the luminosity and is consistent with the data on Faraday rotation measure. The overall broad band spectra are also roughly reproduced if a fraction of energy is allowed to be converted the non-thermal population of electrons. The current results are therefore encouraging, and the further work is in progress.

\ack

The present work was supported by the Polish Grant 1P03D~008~29 and the Polish
Astroparticle Network 621/E-78/SN-0068/2007. 

\bibliography{czerny}

\end{document}